\documentclass[12pt]{article} \topmargin= -0.5cm \textwidth 6.5in
=0cm \headsep =0cm
\usepackage {graphicx}

\begin{document}

\title{Strong magnetic field asymptotic behaviour
 for the fermion-induced  effective energy in the presence of a magnetic flux tube}
\author{Pavlos Pasipoularides  \footnote{paul@central.ntua.gr} \\
       Department of Physics, National Technical University of
       Athens \\ Zografou Campus, 157 80 Athens, Greece}
\date{ }
       \maketitle

\begin{abstract}
In Ref. \cite{3}, we presented an asymptotic formula for the
fermion-induced effective energy in 3+1 dimensions in the presence
of a cylindrically symmetric inhomogeneous strong magnetic field.
However, there are some points which were not clearly explained.
In fact, the arguments, which led us to the asymptotic formula,
are based on a numerical study of the integral of Eq. (10), as we
will see in the main part of this paper. The aim of this work is
to present this study in detail.
\end{abstract}

\section{Introduction}

Many authors have dealt with the numerical study of the
fermion-induced effective energy, in the presence of magnetic
fields of the form of a flux tube, see Refs. \cite{1,2,3,4,5}. An
analytical study of the same problem in two dimensional Euclidean
space can be found in Ref. \cite{2str}.

A similar topic, is the study of the gluon-induced effective
energy in the presence of a colored magnetic flux tube
\cite{9,10}. However, in this case there is an essential
difference: the effective energy has nonzero imaginary part and so
the system is unstable.

It is worth mentioning the recently developed functional method in
Refs. \cite{4,6,7}, which is applicable even for magnetic fields
which do not exhibit a special kind of symmetry. For example, this
method can be applied to a system of two separated magnetic flux
tubes.

In Ref. \cite{3} we presented an asymptotic formula (Eq. (49)) for
the effective energy when the characteristic magnetic field
strength $B_{m}=\int\vec{B}(\vec{x})\cdot d\vec{S} /\pi d^{2}$ (or
the magnetic flux $\Phi$ of the tube) tends to infinity and the
spatial size $d$ of the magnetic flux tube is kept fixed. However,
there are some points which were not clearly explained. In fact,
the arguments, which led us to the asymptotic formula, are based
on a numerical study of the integral of Eq. (10), as we will see
in the main part of this paper. The aim of this work is to present
this study in detail. The final asymptotic formula is given by Eq.
(23).

In addition, in Sec. 4 we compared the asymptotic formula of Eq.
(23) with the strong field limit of the derivative expansion and
find agreement.

\section{Effective energy in the presence of a magnetic flux tube}

The 3+1 dimensional fermion-induced effective energy in the
presence of a magnetic field is given by the equation:
\begin{eqnarray}
E_{eff(3+1)}=-\frac{1}{2iT}Tr\ln(\not\!\!D^{2}+m^{2}_{f})
\end{eqnarray}
where  $\not\!\!\!D=\gamma^{\mu}D_{\mu}$ $(\mu=0,1,2,3)$ and
$D_{\mu}=\partial_{\mu}-ieA_{\mu}$. The gamma matrices satisfy the
relationship $\{\gamma^{\mu},\gamma^{\nu}\}=2g^{\mu\nu}$ and $T$
is the total length of time.

We have shown in Ref. \cite{3} that a magnetic field with strength
independent of $z$ coordinate, directed towards $z$-axis, has
renormalized effective energy per unit length
$L_{z}$\footnote{$L_{z}$ is the length of the space box towards
$z$ direction.} equal to \footnote{For the case of massless
fermions ($m_{f}=0$) the renormalized effective energy is given by
appendix A in Ref. \cite{3}.}
\begin{eqnarray}
E_{eff(3+1)}^{(ren)}=\frac{1}{4\pi}\sum_{\{n\}}(E_{\{n\}}+m_{f}^{2})\ln(\frac{E_{\{n\}}+m_{f}^{2}}{m_{f}^{2}})-\frac{1}{4\pi}\sum_{\{n\}}E_{\{n\}}
\end{eqnarray}
where $E_{\{n\}}$ are the eigenvalues of the planar operator
$(\gamma^{m}D_{m})^{2}$ $(m=1,2)$, and $\{n\}$ is a set of quantum
numbers. Note that $E_{\{n\}}\geq 0$ since the operator
$(\gamma^{m}D_{m})^{2}$ is positive definite as the square of a
hermitian operator. For the sake of simplicity we will drop the
index $(ren)$ in the rest of this paper.

If the magnetic field is cylindrically symmetric with finite
magnetic flux, the spectrum of the operator
$(\gamma^{m}D_{m})^{2}$ is continuous  \footnote{Besides
continuous spectrum, zero modes may exist according to the
Aharonov-Casher theorem \cite{12}.}. Thus we can set
$E_{\{n\}}=k^{2}$.

In order to perform the summation over $\{n\}$ that appears in Eq.
(2), we need a density of states, given by the equation:
\begin{equation}
\rho(k)=\sum_{l,s}\left(\rho_{l,s}^{(free)}(k)+\frac{1}{\pi}\frac{d\delta_{l,s}(k)}{dk}\right)=\sum_{l,s}\rho_{l,s}^{(free)}(k)+\frac{1}{\pi}\frac{d\Delta(k)}{dk}
\end{equation}
where $\sum_{l,s}\rho_{l,s}^{(free)}(k)$ is the density of states
for free space, and $\delta_{l,s}(k)$ is the phase shift which
corresponds to $l^{th}$ partial wave with momentum $k$ and spin
$s$. The function $\Delta(k)$ is defined by the equation:
\begin{equation}
\Delta(k)=\lim_{L\rightarrow+\infty}\sum_{s,l=-L}^{L}\delta_{l,s}(k)
\end{equation}
From Eqs. (2) and (3), if we drop the field independent term
$\sum_{l,s}\rho_{l,s}^{(free)}(k)$ and integrate by parts, we
obtain
\begin{equation}
 E_{eff(3+1)}=-\frac{1}{2\pi^{2}}\int_{0}^{+\infty}k\ln\left(\frac{k^{2}+m^{2}_{f}}{m_{f}^{2}}\right) (\Delta(k)-c) dk
\end{equation}
where $c=\lim_{k \rightarrow +\infty}\Delta(k)$. Our numerical
study, in Ref. \cite{3}, shows that $c=-\pi \phi^{2}$ ($\phi=e
\Phi/2 \pi$). This means that $c$ is independent of the special
form of the magnetic field we examine and depends only on the
total magnetic flux of the field $\Phi$. Also, in Ref. \cite{3},
we see that the function $\Delta(k)$, when $k\rightarrow +\infty$,
tends to $c$ fairly rapidly, and thus, the integral over $k$ in
Eq. (5) is convergent.

Eq. (5) is suitable for the numerical computation of the 3+1
dimensional fermion-induced effective energy in the presence of a
magnetic flux tube. The calculation of the phase shifts is
performed by solving an ordinary differential equation. For
details see Refs. \cite{3,phase}.

\section{The asymptotic behaviour for the effective
 energy when $B_{m}\rightarrow +\infty$ and $d$ and $m_{f}$ are kept fixed}

The function $\Delta(k)$ depends on three quantities: $k$,
$B_{m}=2\phi/d^{2}$ and $d$. From these we construct the following
two dimensionless quantities\footnote{We have assumed the
rescaling $B_{m}\rightarrow eB_{m}$.}: $k B_{m}^{-1/2}$ and
$B_{m}d^{2}$ (=$2\phi$). For dimensional reasons the function
$\Delta(k)$ can be put into the form
$\Delta(k)=G(kB_{m}^{-\frac{1}{2}},B_{m} d^{2})$. Setting
$\Delta(k)=G(kB_{m}^{-\frac{1}{2}},B_{m} d^{2})$ and making the
change of variable $y=kB_{m}^{-\frac{1}{2}}$ in Eq. (5) we obtain
\begin{eqnarray}
E_{eff(3+1)}&=&-\frac{B_{m}}{2\pi^{2}}\int_{0}^{+\infty}y\ln\left(\frac{y^{2}+m^{2}_{f}/B_{m}}{m^{2}_{f}/B_{m}}\right)\left(G(y,B_{m}
d^{2})-c\right)dy\\
&=&-\frac{B_{m}}{2\pi^{2}}\int_{0}^{+\infty}y\left(\ln(y^{2}+m^{2}_{f}/B_{m})-\ln(m^{2}_{f}/B_{m})\right)\left(G(y,B_{m}
d^{2})-c\right)dy
\end{eqnarray}

In the strong magnetic field case ($B_{m}/m_{f}^{2}>>1$) we
obtain:
\begin{eqnarray}
E_{eff(3+1)}\rightarrow
&-&\frac{B_{m}}{\pi^{2}}\int_{0}^{+\infty}y\ln y\left(G(y,B_{m}
d^{2})-c\right)dy \nonumber \\
&-&\frac{B_{m}}{2\pi^{2}}\ln(B_{m}/m^{2}_{f})\int_{0}^{+\infty}y\left(G(y,B_{m}
d^{2})-c\right)dy
\end{eqnarray}
There are many ways to achieve a large ratio $B_{m}/m_{f}^{2}$ by
changing the three independent variables $B_{m}$, $m_{f}$ and $d$.
However, the most interesting case is when $B_{m}\rightarrow
+\infty$ and $m_{f}$ and $d$ are kept fixed (see also Refs.
\cite{1,2,3,4}). This means that we keep the spatial size of the
magnetic field configuration fixed and we increase the
characteristic magnetic field strength $B_{m}$ or the magnetic
flux $\phi=B_{m}d^{2}/2$.

From Eq. (8) we see that the asymptotic behaviour of the effective
energy when $B_{m}\rightarrow +\infty$ ($m_{f}$ and $d$ are kept
fixed) depends on the asymptotic behaviour of the integrals:
\begin{eqnarray}
I_{1}(B_{m}d^{2})&=&\int_{0}^{+\infty}y\left(G(y,B_{m}
d^{2})-c\right)dy \\ I_{2}(B_{m}d^{2})&= & \int_{0}^{+\infty}y\ln
y\left(G(y,B_{m} d^{2})-c\right)dy
\end{eqnarray}
In the case of the integral of Eq. (9) it was possible to find an
analytical expression:
\begin{equation}
\int_{0}^{+\infty} y(G(y,B_{m}d^{2})-c)dy=\frac{1}{12B_{m}}\int
d^{2}\vec{x} B^{2}(\vec{x})
\end{equation}
In order to give a proof for the above equation we will use Eq.
(13) in Ref. \cite{3}
\begin{equation}
Tr(\gamma^{m}D_{m})^{2}=\sum_{\{n\}}E_{\{n\}}=-\frac{1}{6\pi}\int
d^{2}\vec{x} B^{2}(\vec{x})\;
\end{equation}
By using Eq. (3) for the density of states we obtain
\begin{eqnarray}
\sum_{\{n\}}E_{\{n\}}=\int_{0}^{+\infty}
k^{2}\rho(k)dk=\int_{0}^{+\infty}
k^{2}\left(\sum_{l,s}\rho_{l,s}^{(free)}(k)+\frac{1}{\pi}\frac{d\Delta(k)}{dk}\right)dk
\end{eqnarray}
If we drop the field independent term
$\sum_{l,s}\rho_{l,s}^{(free)}(k)$ and set again
$\Delta(k)=G(kB_{m}^{-\frac{1}{2}},B_{m} d^{2})$ and
$y=kB_{m}^{-1/2}$ we find
\begin{eqnarray}
\sum_{\{n\}}E_{\{n\}}&=&\int_{0}^{+\infty}
k^{2}\frac{1}{\pi}\frac{d\Delta(k)}{dk}dk \nonumber \\
&=&-\frac{2}{\pi}\int_{0}^{+\infty} k(\Delta(k)-c)dk \nonumber \\
&=&-\frac{2}{\pi}B_{m}\int_{0}^{+\infty}
y(G(kB_{m}^{-1/2},B_{m}d^{2})-c)dy
\end{eqnarray}
Eq. (11) is a straightforward result of Eqs. (14) and (12).

For convenience we write
\begin{equation}
B(\vec{x})=B_{m}\tilde{B}(\vec{x}/d)
\end{equation}
where the magnetic field $\tilde{B}(\vec{x})$ has a characteristic
magnetic field strength $\tilde{B}_{m}=1$ and range $d=1$, Eq.
(11) can then be written in the form
\begin{equation}
\int_{0}^{+\infty}
y(G(y,B_{m}d^{2})-c)dy=\frac{B_{m}d^{2}}{12}\int d^{2}\vec{x}
\tilde{B}^{2}(\vec{x})
\end{equation}
From the above relationship it is obvious that the integral
$\int_{0}^{+\infty} y(G(y,B_{m}d^{2})-c)dy$ is proportional to
$B_{m}d^{2}$.

\begin{figure}[h]
\begin{center}
\includegraphics[scale=0.5,angle=-90]{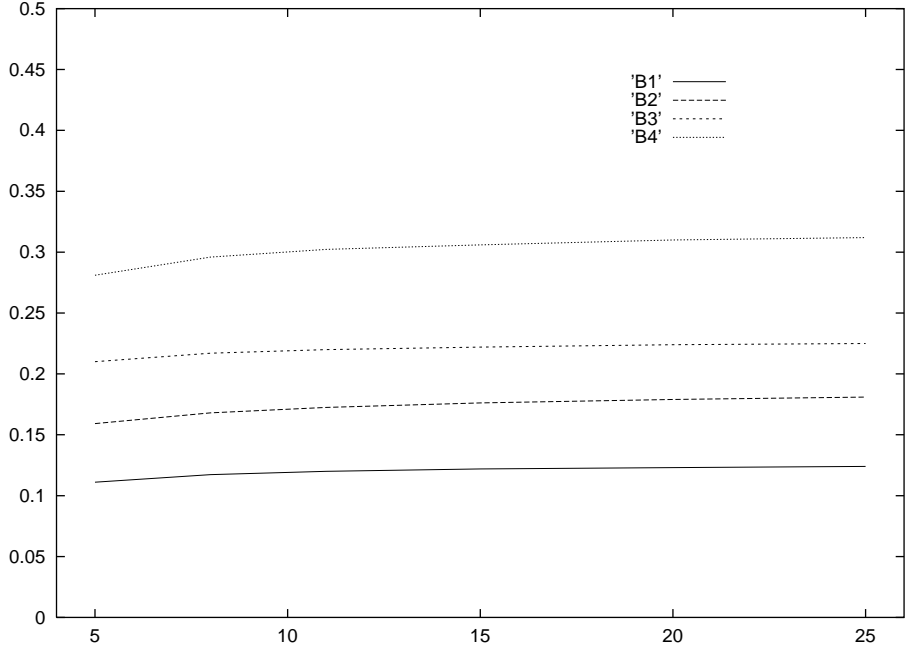}
\end{center}
\caption { $-(1/B_{m}d^{2}) \int_{0}^{\infty}y \ln y
(G(y,B_{m}d^{2})-c) dy $ as a function of $B_{m}d^{2}$ for the
magnetic fields $B_{1}$, $B_{2}$, $B_{3}$, $B_{4}$.} \label{1}
\end{figure}

I could not find an analytical expression, like that of Eq. (16),
for the integral $I_{2}(B_{m}d^{2})$ of Eq. (10). However, a
numerical study, which is presented in Fig. \ref{1}, suggests a
linear \footnote{Our numerical study can not exclude the
possibility of a weak logarithmic growing of
$I_{2}(B_{m}d^{2})/B_{m}d^{2}$, as we will see at the end of this
section} asymptotic behaviour for this integral \footnote{This
implies that the asymptotic behaviour of the integral of Eq. (10)
is the same with that of integral $\int_{0}^{+\infty}
y(G(y,B_{m}d^{2})-c)dy$.}, when $B_{m}\rightarrow +\infty$,
independently of the particular shape of the magnetic field. Thus
we can write
\begin{equation} \int_{0}^{+\infty}y \ln
y \left(G(y,B_{m}d^{2})-c \right) dy \rightarrow c1\: B_{m}d^{2}
\end{equation}
where the constant $c1$ is independent of $B_{m}d^{2}$, and its
value depends only on the special form of the magnetic field
configuration.

As we see in Fig. \ref{1} we have performed numerical calculations
for four magnetic field configurations with quite different kinds
of inhomogeneity.
\begin{eqnarray}
B_{1}(r)&=&B_{m}\: \frac{1}{(r^{2}/d^{2}+1)^{2}}\\
B_{2}(r)&=&B_{m}\: \exp(-\frac{r^{2}}{d^{2}})\\ B_{3}(r)&=&B_{m}\:
\theta(d-r) \\ B_{4}(r)&=&B_{m}\:3(1-\frac{r}{d})\: \theta(d-r)
\end{eqnarray}
We expect that our numerical work, presented in Fig. \ref{1},
convinces the reader that the asymptotic behaviour of
$I_{2}(B_{m}d^{2})$, as given by Eq. (17), does not depend on the
kind of inhomogeneity of the magnetic flux tube.

Thus, when $B_{m}\rightarrow +\infty$ and $m_{f}$ and $d$ are kept
fixed, from Eqs. (8),(16) and (17) we obtain
\begin{eqnarray}
E_{eff(3+1)}\rightarrow
-\frac{c1\:B_{m}^{2}d^{2}}{\pi^{2}}-\frac{B_{m}^{2}d^{2}}{24\pi^{2}}\ln(B_{m}/m^{2}_{f})\int
d^{2}\vec{x} \tilde{B}^{2}(\vec{x})
\end{eqnarray}
We see that the logarithmic term dominates, so we can write
\begin{eqnarray}
E_{eff(3+1)}^{asympt}=-\frac{B_{m}^{2}d^{2}}{24\pi^{2}}\ln(B_{m}/m^{2}_{f})\int
d^{2}\vec{x} \tilde{B}^{2}(\vec{x})
\end{eqnarray}
If, instead of the characteristic magnetic field strength
$B_{m}=\Phi/\pi d^{2}$ we use the parameter $\phi=e\Phi/2 \pi$ we
see that the fermion-induced effective energy, for $\phi
\rightarrow +\infty$, is proportional to $-\phi^{2}\ln\phi$.
\begin{eqnarray}
E_{eff(3+1)}^{asympt} \sim -\phi^{2}\ln\phi
\end{eqnarray}
This asymptotic behaviour is correct, independently of the
particular shape of the magnetic flux tube, as it was shown in the
above numerical study (See Fig. \ref{1}).

Note that the above asymptotic formula, of Eq. (23), is valid and
for other cases of large ratio $B_{m}/m_{f}^{2}$. For example we
note two: a) for $m_{f}\rightarrow 0$ when $B_{m}$ and $d$ are
kept fixed and b) for $d\rightarrow 0$ when $m_{f}$ and
$\phi=B_{m} d^{2}/2$ are kept fixed. For these two cases, the
asymptotic formula, of Eq. (23), is obtained straightforwardly
from Eqs. (8) and (16), and thus knowledge of the asymptotic
behaviour of the integral $I_{2}(B_{m}d^{2})$ is not necessary.

Finally, we note that our numerical study in Fig. \ref{1} is not
sufficient to exclude a logarithmic growing of
$I_{2}(B_{m}d^{2})/B_{m}d^{2}$. However this logarithmic
dependence, even if it really exists, should be weak compared to
the logarithmic dependence which appears in the asymptotic formula
of Eq. (23), otherwise it would be clearly shown in Fig. \ref{1}
and it is not.

In order to observe a possible weak logarithmic growing we ought
to perform computations in an exponentially large range of values
of $B_{m}$. We could not execute computations for so wide a range,
as the differential equation for the computation of the phase
shifts (Eq. (69) in Ref. \cite{3}) is stiff for large $B_{m}$.
However, we tried an estimate of this logarithmic dependence by
fitting a curve of the form $c_{1}+c_{2}\ln(B_{m}d^{2})$ to our
data, as they are presented in Fig. \ref{1}. We find that
$c_{2}=0.0047\pm 0.0005$ for $B_{1}$, $c_{2}=0.0100\pm 0.0005$ for
$B_{2}$, $c_{2}=0.0062\pm 0.0005$ for $B_{3}$ and $c_{2}=0.0120\pm
0.0006$ for $B_{4}$. These results imply that, even if this
logarithmic term $c_{2}\ln(B_{m}d^{2})$ really exists, it would
not contribute to the effective energy (as it is given by the
asymptotic formula of Eq. (23) above) more than 5 per cent.

\section{Derivative expansion for $B_{m}\rightarrow +\infty$ when $d$ and $m_{f}$ are kept fixed}

An approximate way for the computation of the fermion-induced
effective energy is the derivative expansion \cite{der4,der5}.
This method gives accurate results for smooth magnetic field
configurations. Note that the derivative expansion fails for the
magnetic fields of Eqs. (20) and (21), as the magnetic field of
Eq. (20) is discontinuous and the magnetic field of Eq. (21) has
discontinuous first derivative.

Our aim is to compare the strong field limit of the derivative
expansion with the asymptotic formula of Eq. (23).

If we keep the first two terms of the derivative expansion, the
effective energy, in the case of 3+1 dimensions, is given by the
equation
\begin{equation}
E_{eff(3+1)}^{(der)}=E^{(0)}_{(3+1)}[B]+E^{(1)}_{(3+1)}[B,(\partial
B)^{2}]
\end{equation}
where
\begin{eqnarray}
 E_{(3+1)}^{(0)}=\int d^{2}\vec{x}\frac{B^{2}(\vec{x})}{8\pi^{2}}\int_{0}^{+\infty}\frac{1}{s^{2}}(\coth(s)-\frac{1}{s}-\frac{s}{3})
 \: e^{-s \: m_{f}^{2}/B(\vec{x})} ds
\end{eqnarray}
\begin{eqnarray}
E_{(3+1)}^{(1)} =(\frac{1}{8\pi})^{2} \int d^{2}\vec{x}
\frac{(\nabla B)^{2}}{B(\vec{x})}\int_{0}^{+\infty}\frac{1}{s}
 (\frac{d}{ds})^{3}[s\coth(s)] \:e^{-s \:m_{f}^{2}/B(\vec{x})} ds
\end{eqnarray}

For $B_{m}\rightarrow +\infty$ from Eqs. (26) and (27) we obtain
\begin{eqnarray}
 E_{(3+1)}^{(0)}\rightarrow -\frac{B_{m}^{2}d^{2}}{24\pi^{2}}\ln(B_{m}/m^{2}_{f})\int
d^{2}\vec{x} \tilde{B}^{2}(\vec{x})
\end{eqnarray}
\begin{eqnarray}
E_{(3+1)}^{(1)} \rightarrow B_{m}(\frac{1}{8\pi})^{2} 24
\zeta'(-2) \int d^{2}\vec{x} \frac{(\nabla
\tilde{B})^{2}}{\tilde{B}(\vec{x})}
\end{eqnarray}
Eq. (29) can be derived straight from Eq. (27) if we assume that
$B_{m}>>m_{f}^{2}$ (for details see Ref. \cite{der5}). In order to
derive Eq. (28) from Eq. (26) we have used a similar way with that
presented in Ref. \cite{strong}.

Thus Eq. (28), which gives the strong field limit of the
derivative expansion, is in agreement with the asymptotic formula
of Eq. (23).

\section{Conclusions}
We emphasize that the arguments, which led us to the asymptotic
formula of Eq. (23), are based on a numerical study. In
particular, we computed numerically $I_{2}(B_{m}d^{2})$ for four
magnetic field configurations with quite different kinds of
inhomogeneity (see Eqs. (18), (19), (20) and (21)). Our results in
Fig. \ref{1} suggest that $I_{2}(B_{m}d^{2})$ is proportional to
$B_{m}d^{2}$ (for $B_{m}d^{2}>>1$) independently of the specific
form of the magnetic field configurations. The consequence of this
result is that also the final formula, which states that
$E_{eff(3+1)}^{asympt} \sim -\phi^{2}\ln\phi$ for $\phi\rightarrow
+\infty$, is valid independently of the specific form of the
magnetic field.

Finally we showed, comparing with the asymptotic formula of Eq.
(23), that the leading term of the derivative expansion gives the
same asymptotic behaviour for the fermion-induced effective energy
in the strong magnetic field regime even in the case (like that of
the magnetic fields of Eqs. (20) and $(21)$) when the derivative
expansion is invalid, because the next to the leading term
diverges.

\section{Acknowledgements}
I am grateful to Professor G. Tiktopoulos for numerous valuable
discussions, as well as Professors K. Farakos and G. Koutsoumbas
for useful comments and suggestions.


\begin{thebibliography}{99}

\bibitem{1} M. Bordag and K. Kirsten, \textit{The ground state
energy of a spinor field in the background of finite radius flux
tube}, Phys. Rev. D 60, 105019 (1999).

\bibitem{2} M. Scandura, \textit{Vacuum energy in the
presence of a magnetic string with delta function profile},
Phys.Rev. D 62, 085024 (2000).

\bibitem{3} P. Pasipoularides, \textit{Fermion-induced effective action in
 the presence of a static inhomogeneous magnetic field}, Phys. Rev.
 D 64, 105011 (2001).

\bibitem{4}  K. Langfeld, L. Moyaerts,  H. Gies, \textit{Fermion-indused quantum action of vortex
system}, Nucl.Phys., B646:158-180 (2002).

\bibitem{5} I. Drozdov, \textit{Vacuum polarization by a magnetic flux of special rectangular
form}, hep-th/0210282.

\bibitem{2str} M. P. Fry, \textit{Fermion determinant for general background gauge
fields}, hep-th/0301097

\bibitem{6} H. Gies, K. Langfeld, \textit{Loops and loop clouds-A numerical approach
to the worldline formalism in QED}, Int.J.Mod.Phys., A17:966-978
(2002).

\bibitem{7} H. Gies, K. Langfeld, \textit{Quantum diffusion of magnetic fields in numerical worldline
approach}, Nucl.Phys., B613:353-365 (2001) .

\bibitem{9} D. Diakonov, M. Maul, \textit{Center-vortex solutions of the Yang-Mills
 effective action in three and four dimensions}, hep-lat/0204012.

\bibitem{10}  M. Bordag, \textit{On the vacuum energy of a color magnetic
vortex}, hep-th/0211080.

\bibitem{phase} E. Farhy, N. Graham, P. Haagensen                                                                                                                                                                                                      and
R.L. Jaffe, \textit{Finite Quantum Fluctuations About Static Field
Configurations}, Phys.Lett. 427B, 334 (1998).

\bibitem{der4} G. Dunne, \textit{An all-orders Derivative Expansion}, Intern. Journ. of Mod. Phys. A, Vol 12, No
6 1143 (1997).

\bibitem{der5} V.P. Gusynin and I.A. Shovkovy, \textit{Derivative
Expansion of the Effective Action for QED in $2+1$ and $3+1$
dimensions}, J. Math. Phys. 40, 5406 (1999).

\bibitem{12}  Y. Aharonov and A. Casher, \textit{Ground state of a spin-$\frac{1}{2}$ charged particle in a two-dimensional magnetic
field}, Phys. Rev. A 19, 2461 (1979).


\bibitem{strong} W. Dittrich, W. Tsai, K. H. Zimmermann, \textit{Evaluation of the effective potential
 in quantum electrodynamics}, Phys. Rev. D19, 2929 (1979).


\end{thebibliography}
\end{document}